\documentclass[12pt]{article}

\begin{document}

\title{Coherent states with elliptical polarization}
\author{E. Colavita and S. Hacyan}

\renewcommand{\theequation}{\arabic{section}.\arabic{equation}}

\maketitle
\begin{center}

{\it  Instituto de F\'{\i}sica,} {\it Universidad Nacional
Aut\'onoma de M\'exico,}

{\it Apdo. Postal 20-364, M\'exico D. F. 01000, Mexico.}

e-mail: hacyan@fisica.unam.mx

\end{center}
\vskip0.5cm

Coherent states of the two dimensional harmonic oscillator are
constructed as superpositions of energy and angular momentum
eigenstates. It is shown that these states are Gaussian
wave-packets moving along a classical trajectory, with a well
defined elliptical polarization. They are coherent correlated
states with respect to the usual cartesian position and momentum
operators. A set of creation and annihilation operators is defined
in polar coordinates, and it is shown that these same states are
precisely coherent states with respect to such operators.

\bigskip


\bigskip


\newpage

\section{Introduction}

Coherent states were introduced by Glauber \cite{glauber} in
quantum optics and have a very wide range of application in
quantum physics. They can be defined in several equivalent ways:
as Gaussian wave packets evolving without spreading, and also as
close analogues of classical states satisfying the minimum
dispersion relation allowed by the Heisenberg uncertainty
principle. One of the various generalizations of coherent states
are the coherent correlated states introduced by Dodonov {\it et
al.}\cite{ccs}, that satisfy the minimum Robertson-Schr\"{o}dinger
uncertainty relation.

A coherent state can be expressed as an infinite superposition of
the number states of a harmonic oscillator. The generalization to
two or more modes can be easily achieved by coupling several
linear harmonic oscillator. This is the usual approach used in
this kind of problems; for instance, Arickx {\it et al.}
\cite{arick} have studied the evolution of Gaussian wave packets
in two and three dimensions in cartesian coordinates.

However, polar coordinates are the natural coordinates for a
two-dimensional oscillator with a single frequency, permitting to
separate directly the angular momentum. Accordingly, it would be
advantageous to have a definition of coherent states that takes
into account the rotational symmetry of the problem. This is
particularly relevant due the recent interest in the angular
momentum of light, associated to electromagnetic fields with axial
symmetry such as Bessel and Laguerre beams (see e. g. Padgett {\it
et al.} \cite{pt}).

In this article, we propose a new kind of coherent states that are
superpositions of states with well defined energy and angular
momentum, and that can be directly related to axially symmetric
fields. The analysis of these states is presented in Section 2,
where it is also shown that they are coherent correlated states
with respect to cartesian position and momentum, and correspond to
two independent elliptical polarizations, rotating clockwise or
counterclockwise. In Section 3, a set of creation and annihilation
operators in polar coordinates are defined, and it is shown that
the states are truly coherent with respect to these operators
because they have minimum dispersion. Some numerical examples are
presented.

\section{Coherent correlated states}

Consider a two dimensional harmonic oscillator with mass $M$ and
frequency $\omega$. Choosing $M$, $\omega^{-1}$, and
$(\hbar/m\omega)^{1/2}$ as units of mass, time and length
respectively, the time dependent Schr\"{o}dinger equation in
cylindrical coordinates takes the form:
\begin{equation}
i \frac{\partial}{\partial t}\Psi (r, \phi, t)= \frac{1}{2}\Big(-
\frac{1}{r} \frac{\partial}{\partial r} r \frac{\partial}{\partial
r}-\frac{1}{r^2}\frac{\partial^2}{\partial \phi^2}+r^2 \Big) \Psi
(r, \phi, t).\label{1}
\end{equation}
The eigenstates of energy and angular momentum are $|n,\pm
l\rangle$ (we take $l\geq 0$). In coordinates representation,
these states are given by the standard text-book solutions (see,
e. g., \cite{frank}):
\begin{equation}
\langle r, \phi|n,\pm l\rangle = \Big( \frac{n!}{\pi (n+l)!}
\Big)^{1/2} e^{\pm i l \phi}e^{-r^2/2}r^{l} L^{l}_n (r^2),
\end{equation}
where $L^l_n (r^2)$ are Laguerre polynomials. The energy of these
states is $E_{n,l}= 2n +l+1 $.

In analogy with the usual definition of coherent states, let us
now define the following states:
\begin{equation}
|\alpha, \beta \rangle_{\pm}= N(\alpha, \beta)
\sum_{n,l=0}^{\infty}\Big( \frac{(n+l)!}{n!}
\Big)^{1/2}\frac{1}{l!}\alpha^n  \beta^l |n,\pm l\rangle,
\end{equation}
where $N(\alpha, \beta)$ is a normalization constant. Clearly, the
evolution of these states is given by
\begin{equation}
e^{-i H } |\alpha, \beta \rangle_{\pm} = |\alpha e^{- 2i t}, \beta
e^{-i t} \rangle_{\pm}.
\end{equation}

The orthogonality condition determines $N(\alpha, \beta)$, that
is:
\begin{equation}
_{\pm}\langle \alpha, \beta | \alpha, \beta \rangle_{\pm}= |
N(\alpha, \beta)|^2 \sum_{n,l=0}^{\infty} \frac{(n+l)!}{n!
(l!)^2}|\alpha|^{2n} |\beta |^{2l} = 1.
\end{equation}
Using the formula
\begin{equation}
\sum_{n=0}^{\infty} \frac{(n+l)!}{n!}x^n = \frac{l!}{(1-x)^{l+1}},
\end{equation}
it follows that
\begin{equation}
| N(\alpha, \beta)| =\sqrt{1- |\alpha|^2}
\exp\Big\{{-\frac{|\beta|^2}{2(1- |\alpha|^2)}}\Big\},
\end{equation}
with the condition $|\alpha|< 1$.

It can also be seen that
$$
_{\pm}\langle \alpha^{\prime}, \beta^{\prime} | \alpha, \beta
\rangle_{\pm}= \frac{\sqrt{1- |\alpha|^2} \sqrt{1-
|\alpha^{\prime}|^2}}{1-\alpha \alpha^{*\prime}}
$$
\begin{equation}
\times \exp\Big\{- \frac{|\beta|^2}{2(1- |\alpha|^2)} -
\frac{|\beta^{\prime}|^2}{2(1- |\alpha^{\prime}|^2)} + \frac{\beta
\beta^{*\prime}}{1- \alpha \alpha^{*\prime}} \Big\},
\end{equation}
and
$$
_{\mp}\langle \alpha^{\prime}, \beta^{\prime} | \alpha, \beta
\rangle_{\pm}= \frac{\sqrt{1- |\alpha|^2} \sqrt{1-
|\alpha^{\prime}|^2}}{1-\alpha \alpha^{*\prime}}
$$
\begin{equation}
\times \exp\Big\{- \frac{|\beta|^2}{2(1- |\alpha|^2)} -
\frac{|\beta^{\prime}|^2}{2(1- |\alpha^{\prime}|^2)} \Big\}.
\end{equation}

Let us now look for the explicit form of the states $|\alpha,
\beta \rangle_{\pm}$ in coordinates representation. Since
\begin{equation}
\langle r, \phi |\alpha, \beta \rangle_{\pm}= \frac{1}{\pi^{1/2}}
N(\alpha, \beta) \sum_{n,l=0}^{\infty} \frac{1}{l!}\alpha^n
\beta^l e^{il\phi}e^{-r^2 /2}r^l L_n^l (r^2),
\end{equation}
and using the generating function formula:
\begin{equation}
\sum_{n=0}^{\infty} a^n L_n^l (x) = (1-a)^{-l-1} e^{xa/(a-1)},
\end{equation}
it follows that
\begin{equation}
\langle r, \phi |\alpha, \beta \rangle_{\pm}=
\frac{N_{\pm}(\alpha, \beta)}{\pi^{1/2}(1-\alpha)} ~ e^{-(1+
\alpha)r^2/2(1- \alpha)} \sum_{l=0}^{\infty} \frac{1}{l!} ~ \Big(
\frac{\beta e^{\pm i \phi}}{1 - \alpha} r \Big)^l .
\end{equation}
Accordingly:
$$
\Psi_{\alpha \beta}^{\pm}(r, \phi) \equiv  \langle r, \phi
|\alpha, \beta \rangle_{\pm}
$$
\begin{equation}
= \frac{1}{\pi^{1/2}} \sqrt{\frac{1+\alpha}{1-\alpha}} ~
\exp\Big\{-\frac{1+ \alpha}{1- \alpha}~\frac{~r^2}{2} +
\frac{\beta e^{\pm i \phi}}{1- \alpha}~r -\frac{|\beta|^2}{2(1-
|\alpha|^2)} \Big\} .\label{213}
\end{equation}

Now, writing the exponent in this last equation in the form:
\begin{equation}
-\frac{1+ \alpha}{1- \alpha}~\frac{~r^2}{2} + \frac{\beta e^{\pm i
\phi}}{1- \alpha}~r = - \frac{1}{2} ({\bf x}-{\bf q}) ({\bf
U}^{-1}+i{\bf V}) ({\bf x}-{\bf q})+ i{\bf p \cdot x},
\end{equation}
we can compare our results with those of Arick {\it et al}
\cite{arick}. Remembering that $\alpha (t) = \alpha e^{- 2 i t}$
and $\beta (t) = \beta e^{- i t}$ (taking $\alpha$ and $\beta$
real in the rest of this section without loss of generality), it
follows that the position ${\bf q}$ and momentum ${\bf p}$ of the
wave-packet center are:
\begin{equation}
{\bf q} = \Big\{ \frac{\beta \cos t}{1+\alpha}, - \frac{\beta \sin
t}{1-\alpha} \Big\},
\end{equation}
\begin{equation}
{\bf p} = \Big\{ \frac{\beta \sin t}{1+\alpha}
\Big(1+\frac{2\alpha \cos^2 t}{|1-\alpha e^{-2it}|^2} \Big)~,~
\frac{\beta \cos t}{1-\alpha}\Big(1-\frac{2\alpha \sin^2
t}{|1-\alpha e^{-2it}|^2}\Big)\Big\}.
\end{equation}
Similarly, we find:
\begin{equation}
{\bf U} = \frac{|1-\alpha e^{2it}|^2}{1-\alpha^2} {\bf 1}
\end{equation}
and
\begin{equation} {\bf V} = \frac{2 \alpha \sin 2t}{|1-\alpha
e^{-2it}|^2} {\bf 1};
\end{equation}
these matrices determine the probability density, $$\rho(x) =
\exp\{-({\bf x}-{\bf q}) {\bf U}^{-1} ({\bf x}-{\bf q})\}$$ and
the current density $$j(x)=\rho(x) [{\bf p} + {\bf V}({\bf x}-{\bf
q})]$$ (see Ref. \cite{arick}).

The above relations describe a Gaussian wave-packet with a
periodically varying variance, and a center moving along the
classical trajectory of a harmonic oscillator. This trajectory is
an ellipse of eccentricity
$$2\frac{\sqrt{\alpha}}{1+\alpha},$$
and with semi-major and semi-minor axis determined by the prameter
$\beta$ :
$$\frac{\beta}{1\pm \alpha}~.$$

Following the formulas of Ref.\cite{arick}, we can readily
calculate the dispersions:
\begin{equation}
\langle \Delta x^2 \rangle = \langle \Delta y^2 \rangle =
\frac{1}{2} ~ \frac{|1-\alpha e^{-2it}|^2}{1-\alpha ^2} ,
\end{equation}
which show that the width of the Gaussian wave-packet oscillates
sinusoidally between the values $\frac{1}{2}(1-\alpha)/(1+\alpha)$
and $\frac{1}{2}(1+\alpha)/(1-\alpha)$. Similarly,
\begin{equation}
\langle \Delta p_x^2 \rangle = \langle \Delta p_y^2 \rangle =
\frac{1}{1-\alpha^2} \frac{2 \alpha^2 \sin^2 2t}{|1-\alpha
e^{-2it}|^2} + \frac{1}{2} ~ \frac{1-|\alpha |^2}{|1-\alpha
e^{-2it}|^2} ,
\end{equation}
and
\begin{equation}
\langle \Delta x \Delta p_x \rangle = \langle \Delta y \Delta p_y
\rangle =  \frac{\alpha \sin 2t}{1-\alpha^2} .
\end{equation}
From these formulas, we notice that:
\begin{equation}
\langle \Delta x^2 \rangle \langle \Delta p_x^2 \rangle - \langle
\Delta x \Delta p_x \rangle^2 =\langle \Delta y^2 \rangle \langle
\Delta p_y^2 \rangle - \langle \Delta y \Delta p_y \rangle^2 =
\frac{1}{4} ~~,
\end{equation}
which is precisely the definition of a coherent correlated state
\cite{ccs}.

\section{Polar operators and coherent states}

In order to compare with the standard results of the
two-dimensional harmonic oscillator, we consider the usual
creation and annihilation operators in Cartesian coordinates:
$$
a_i = \frac{1}{\sqrt{2}}(x_i + \frac{\partial}{\partial x_i })
$$
\begin{equation}
a_i^{\dagger} = \frac{1}{\sqrt{2}}(x_i - \frac{\partial}{\partial
x_i }),
\end{equation}
with $(x_1 , x_2 ) = (x , y) = (r \cos \phi , r \sin \phi )$. They
satisfy the commutation relations:
$$[a_i~, ~a_j^{\dagger}] = \delta_{ij}.$$

Define now the operators
\begin{equation}
A = \frac{1}{\sqrt{2}} ( a_1 + i a_2 )
\end{equation}
\begin{equation}
B = \frac{1}{\sqrt{2}} ( a_1 - i a_2 ).
\end{equation}
Their explicit forms in polar coordinates are:
$$
A =\frac{1}{2} e^{i\phi} \Big(r+\frac{\partial}{\partial r} +
\frac{i}{r}\frac{\partial}{\partial \phi}\Big)
$$
$$
B =\frac{1}{2} e^{-i\phi} \Big(r+\frac{\partial}{\partial r} -
\frac{i}{r}\frac{\partial}{\partial \phi}\Big)
$$
$$
A^{\dagger} =\frac{1}{2} e^{-i\phi}
\Big(r-\frac{\partial}{\partial r} +
\frac{i}{r}\frac{\partial}{\partial \phi}\Big)
$$
\begin{equation}
B^{\dagger} =\frac{1}{2} e^{i\phi} \Big(r-\frac{\partial}{\partial
r} - \frac{i}{r}\frac{\partial}{\partial \phi}\Big),\label{op}
\end{equation}
and they satisfy the commutation relations:
$$[A,~B] = 0 = [A,~B^{\dagger}]$$
\begin{equation}
[A,~A^{\dagger}] = 1 = [B,~B^{\dagger}].
\end{equation}

If we now express the wave functions in coordinate representations
as in Eq. (\ref{213}), then it is easy to see that:
\begin{equation}
e^{i\phi} \Big(\frac{\partial}{\partial r} +
\frac{i}{r}\frac{\partial}{\partial \phi}\Big)~ \Psi_{\alpha
\beta}^{+}(r, \phi) = -\frac{1+\alpha}{1-\alpha}e^{i\phi} r
\Psi_{\alpha \beta}^{+}(r, \phi)
\end{equation}
and
\begin{equation}
e^{-i\phi} \Big(\frac{\partial}{\partial r} -
\frac{i}{r}\frac{\partial}{\partial \phi}\Big)~ \Psi_{\alpha
\beta}^{+}(r, \phi) = \Big[ -\frac{1+\alpha}{1-\alpha}e^{-i\phi} r
+ \frac{2 \beta}{1-\alpha} \Big] \Psi_{\alpha \beta}^{+}(r, \phi).
\end{equation}

These last two relations can be written in a coordinate
independent form in terms of the operators defined in Eqs.
(\ref{op}). Thus, for the righthanded polarized states:
\begin{eqnarray}
(A+\alpha B^{\dagger}) |\alpha \beta \rangle_+  &=& 0 \\
(\alpha A^{\dagger} + B) |\alpha \beta \rangle_+  &=& \beta
|\alpha \beta \rangle_+
\end{eqnarray}

From these expression it is easy to find the expectation values of
the $A$ and $B$ operators in the righthanded states:
$$
_+\langle A \rangle_+ = -\frac{\alpha \beta^* }{1 -|\alpha|^2}
$$
\begin{equation}
_+\langle B \rangle_+ = \frac{\beta }{1 -|\alpha|^2}.
\end{equation}

In a similar way, it can be seen that for the lefthanded polarized
states:
\begin{eqnarray}
(\alpha A^{\dagger}+ B) |\alpha \beta \rangle_-  &=& 0 \\
(A + \alpha B^{\dagger}) |\alpha \beta \rangle_-  &=& \beta
|\alpha \beta \rangle_-
\end{eqnarray}
and the expectations values in these states are:
$$
_-\langle A \rangle_- = \frac{\beta }{1 -|\alpha|^2}
$$
\begin{equation}
_-\langle B \rangle_- = -\frac{\alpha \beta^*}{1 -|\alpha|^2}.
\end{equation}
Notice that the eigenvalue equations for the $|\alpha \beta
\rangle_-$ are the same for the $|\alpha \beta \rangle_+$ states
if the interchange $A\leftrightarrow B$ is made.

As for the expectation values of quadratic operators, some lengthy
but straightforward calculations  show that:
\begin{equation}
_+\langle A^{\dagger}A \rangle_+ = \frac{|\alpha|^2 |\beta|^2 }{(1
-|\alpha|^2)^2}~,
\end{equation}
\begin{equation}
_+\langle B^{\dagger}B \rangle_+ = \frac{|\beta|^2 }{(1
-|\alpha|^2)^2}~,
\end{equation}
\begin{equation}
_+\langle A^{\dagger}B \rangle_+ = -\frac{\alpha^* \beta^2 }{(1
-|\alpha|^2)^2}~,
\end{equation}
\begin{equation}
_+\langle A^2 \rangle_+ = \frac{\alpha^2 \beta^{*2} }{(1
-|\alpha|^2)^2}~,
\end{equation}
\begin{equation}
_+\langle B^2 \rangle_+ = \frac{\beta^2}{(1 -|\alpha|^2)^2}~,
\end{equation}
\begin{equation}
_+\langle A B \rangle_+ = -\frac{\alpha |\beta|^2 }{(1
-|\alpha|^2)^2}~.
\end{equation}

The equivalent expressions for the expectation values in the
left-handed state can be obtained directly by the interchange
$A\leftrightarrow B$ combined with $_+\langle ... \rangle_+
\leftrightarrow _-\langle ... \rangle_-$.

Defining the Hermitian operators $Q_{A,B}$ and $P_{A,B}$ as
${A,B}=\frac{1}{2}(Q_{A,B}+iP_{A,B})$, it follows from the above
formulas that
\begin{equation}
_{\pm}\langle (\Delta Q_{A,B})^2 \rangle_{\pm} = _{\pm}\langle
(\Delta P_{A,B})^2 \rangle_{\pm} ~= 1.
\end{equation}
Therefore, the states under consideration are indeed coherent
states, that is, they have minimum dispersion with respect to the
$A$ and $B$ operators.

 Finally, we recall that the operators $a_i$ and
$a^{\dagger}_i$ also generate an U(2) algebra (see, e. g.,
\cite{frank}). In terms of the standard operators $a_i$ and
$a^{\dagger}_i$ and the new operators $A$ and $B$, the components
of an angular momentum operator ${\bf J}$ can be defined as:
$$
J_x + i J_y \equiv \frac{1}{2}[(a_x^{\dagger} + i a_y^{\dagger})
(a_x + i a_y)]
$$
\begin{equation}
=  B^{\dagger} A = \frac{1}{4}e^{2i \phi} \Big[r^2 + \frac{1}{r}
\frac{\partial}{\partial r} - \frac{\partial^2}{\partial
r^2}+\frac{1}{r^2}\frac{\partial^2}{\partial  \phi^2} +
2i\Big(\frac{1}{r^2}\frac{\partial}{\partial \phi}
-\frac{1}{r}\frac{\partial^2}{\partial \phi \partial r}\Big)\Big]
,
\end{equation}
$$
J_z \equiv \frac{i}{2}(a_y^{\dagger}a_x - a_x^{\dagger} a_y)
$$
\begin{equation}
= -~\frac{1}{2} ~(A A^{\dagger} - B B^{\dagger})= \frac{i}{2}
\frac{\partial}{\partial \phi}~~,
\end{equation}
and the number operator is
\begin{equation}
N = a_x^{\dagger} a_x + a_y^{\dagger} a_y = A^{\dagger} A +
B^{\dagger} B ,
\end{equation}
which is simply related to the Hamiltonian appearing in Eq.
(\ref{1}) : $H = N + 1$.

\section{Results and conclusions}

We have performed a numerical calculations in order to illustrate
the evolution of the states described in this paper. In Fig. 1, we
present a state corresponding to left handed polarization with
particular values of the parameters $\alpha$ and $\beta$. As
expected, the Gaussian wave packet rotates along a ellipse and
changes its dispersion periodically; the form of the wave-packet
is recovered after each cycle, in accordance with the results
obtained by Arickx {\it et al.} \cite{arick}. A cross section of
the $|\Psi_{\alpha \beta}^-(r, \phi) |^2 $ is shown in the figure.

An interesting feature of the states we have defined in this
article is that they are coherent correlated states with respect
to the cartesian position and momentum operators. However, a new
set of position and momentum operators can be defined in polar
coordinates, and these same states turn out to be linear
combinations of the eigenstates of these operators, with the
peculiarity of being truly coherent with respect to these
operators, having the minimum dispersion allowed by the Heisenberg
uncertainty principle.

In conclusion, the use of polar coordinates permits to see some
features that are not evident in cartesian coordinates. The
relation between classical and quantum motions is made explicit
including  the rotational degrees of freedom that correspond to
different polarizations.

\section*{Acknowledgments}
We are grateful to R. J\'{a}uregui for many fruitful comments and
discussions.

\bigskip
\section*{Figure caption}

\bigskip
\noindent Fig. 1- An horizontal cross section of the square
modulus $|\Psi_{\alpha \beta}^{\pm}(r, \phi)|^2$ with parameters
$\alpha=0.4$ and $\beta=5$, in two-dimensional coordinates space.
The cut is at $|\Psi_{\alpha \beta}^{\pm}(r, \phi)|^2 = 0.05$. The
classical trajectory, an ellipse with eccentricity $2\sqrt{\alpha}
/(1+\alpha)$, is added to illustrate the corresponding classical
motion.

\end{document}